\documentclass[preprint,12pt]{elsarticle}
\usepackage{graphicx}
\usepackage{amssymb}
\usepackage{amsmath}
\usepackage{subfig}
\usepackage{multirow}
\usepackage{colortbl,color}
\usepackage{algorithm}
\usepackage[page,title,titletoc,header]{appendix}
\usepackage{multirow}
\usepackage{amsthm}
\usepackage{hyperref}
\usepackage{bm}
\biboptions{comma,sort&compress}
\journal{ }

\begin{document}

\begin{frontmatter}
\title{Lattice Boltzmann modelling of intrinsic permeability}
\author[KFUPM]{Jun~Li}
\address[KFUPM]{Center for Integrative Petroleum Research, \\ College of Petroleum Engineering and Geosciences, \\ King Fahd University of Petroleum $\&$ Minerals, Saudi Arabia}

\author[Strath]{Minh Tuan ~Ho}
\author[Strath]{Lei~Wu}
\author[Strath]{Yonghao~Zhang}
\address[Strath]{James Weir Fluids Laboratory, \\ Department of Mechanical Engineering \& Aerospace Engineering, \\University of Strathclyde, Glasgow, UK}
\begin{abstract}
Lattice Boltzmann method (LBM) has been applied to predict flow properties of porous media including intrinsic permeability, where it is implicitly assumed that the LBM is equivalent to the incompressible (or near incompressible) Navier-Stokes equation. However, in LBM simulations, high-order moments, which are completely neglected in the Navier-Stokes equation, are still available through particle distribution functions. To ensure that the LBM simulation is correctly working at the Navier-Stokes hydrodynamic level, the high-order moments have to be negligible. This requires that the Knudsen number ({\it Kn}) to be small so that rarefaction effect can be ignored. In this technical note, we elaborate this issue in LBM modelling of porous media flows, which is particularly important for gas flows in ultra-tight media.
\end{abstract}
\end{frontmatter}

\section{Introduction}\label{s:Introduction}

Lattice Boltzmann method (LBM) is a popular method for calculation of flow properties of porous media e.g. intrinsic permeability \cite{Pan2006}-\cite{Prestininzi2016}. The standard LBM is regarded as an alternative method to computational fluid dynamics (CFD), equivalent to solving the incompressible (or near incompressible) Navier-Stokes equation. Through the Chapman-Enskog expansion,  we can show that the convergence of LBM to the incompressible Navier-Stokes equation in the low Mach and Knudsen numbers limits. However, these two methods are very different. For example, the third-order and higher-order moments are completely neglected in the isothermal Navier-Stokes equation while they are still $available$ in LBM simulations through particle distribution functions, despite that they can be negligibly small when the Knudsen number ({\it Kn}) is close to zero. Therefore, the high-order moments have to be negligible if the LBM simulation is correctly working at the Navier-Stokes level, which has been commonly oversighted in simulating flows in porous media.

In LBM simulations, the model parameters  can be correlated with the kinematic viscosity $\nu$ as follows:

\begin{equation} \label{eq:nu}
\begin{aligned}
   \nu=\dfrac{(\tau-0.5)c\Delta x}{3},
\end{aligned}
\end{equation}
where $\Delta x, \Delta t$ and $\tau$ are the grid or lattice length, the time step, and the normalized relaxation time, respectively; and $c=\Delta x/\Delta t.$ Therefore, we have flexibility in selecting $c, \Delta x, \tau$ for any given physical $\nu$. In the following examples, we will demonstrate that choice of these model parameters may lead to finite $Kn$, which incurs unintentional rarefaction effect at the Navier-Stokes level simulations.

\section{Results and Discussions}\label{s:Results}

\begin{figure}[H]
	\centering
	\includegraphics[width=0.6\textwidth]{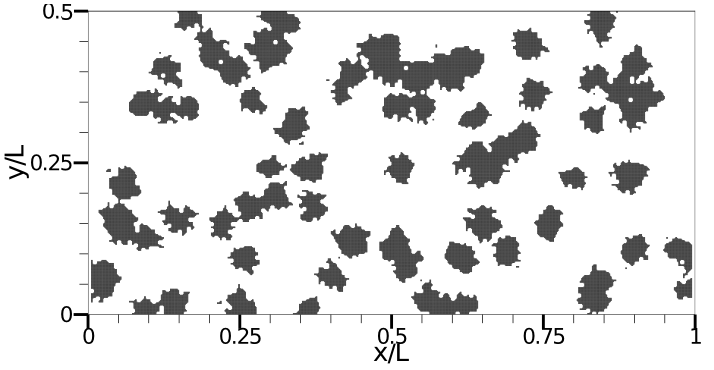}
	\caption{The 2D geometry used in the pore-scale simulations, where the symmetric boundary condition is used at the top and bottom surfaces, while the pressure boundary condition is applied at the inlet and outlet.}
	\label{fig2d:geometry}
\end{figure}

The intrinsic permeability depends on pore structure rather than the flow properties. First, we apply the D2Q9 LBM model to simulate a pressure-driven flow along the $x$ direction in a 2D randomly generated quartet structure, see Fig. \ref{fig2d:geometry}. The applied pressure difference at the inlet and outlet is so small that the flows are in the Stokes flow regime. The permeability is then calculated according to the Darcy law with simulated mass flow rate. The porosity is 0.7394 and the resolutions are $N_x*N_y=400*200$, where $N_x$ and $N_y$ are the lattice numbers in the $x$ and $y$ directions. We also increase the resolutions to 800*400, and 1200*600 respectively. The porosity is also slightly changed to 0.7532 and 0.7578 respectively.

\begin{figure}[H]
	\centering
	\includegraphics[width=0.6\textwidth]{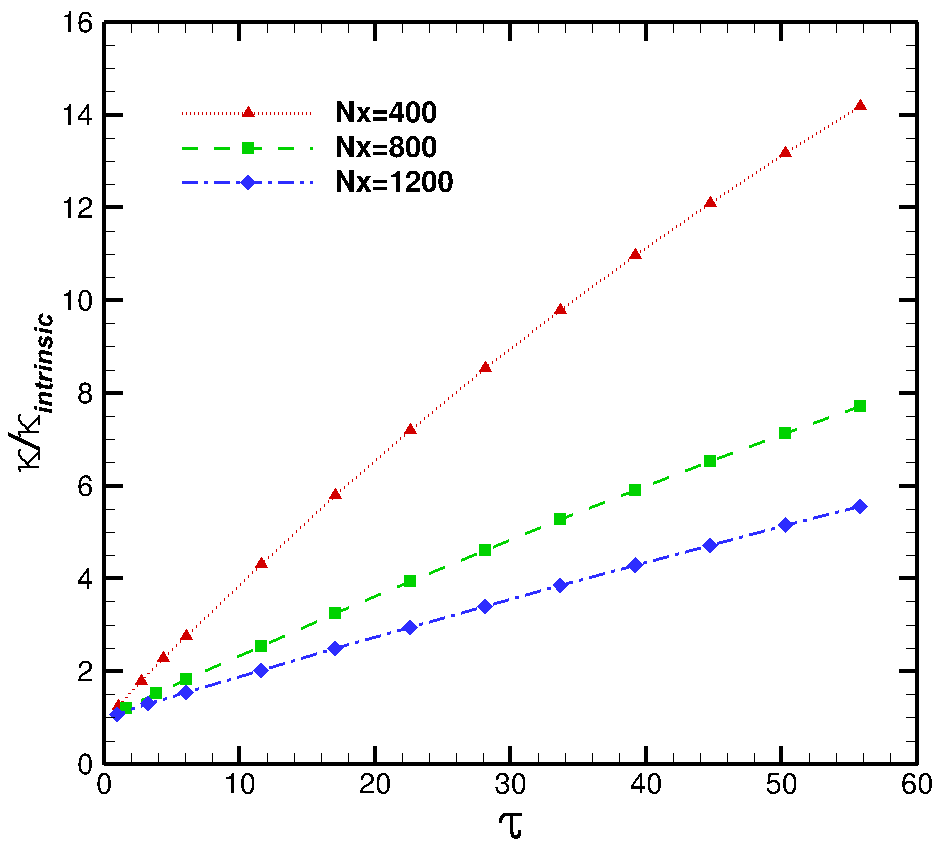}
	\caption{The simulated permeability normalized by the intrinsic permeability against the relaxation time $\tau$ for 3 different resolutions.}
	\label{fig2d:tau}
\end{figure}

\begin{figure}[H]
	\centering
	\includegraphics[width=0.6\textwidth]{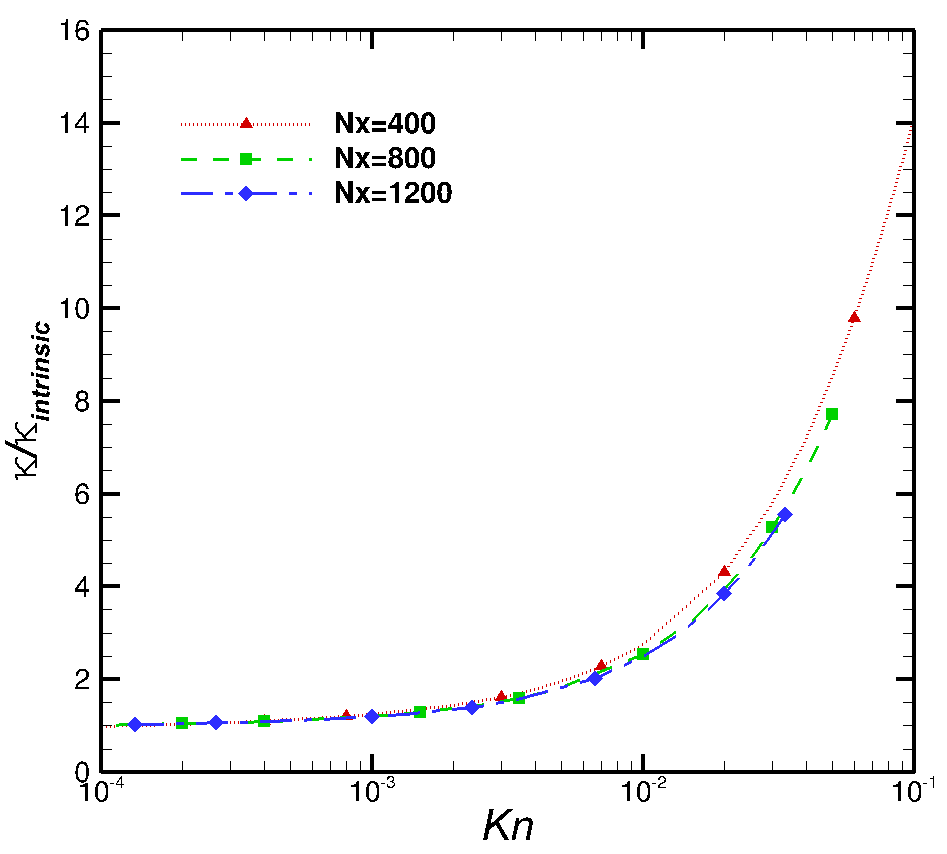}
	\caption{The simulated permeability normalized by the intrinsic permeability against $Kn$.}
	\label{fig2d:Kn}
\end{figure}

Fig.~\ref{fig2d:tau} shows that the permeabilities for 3 resolutions are very different even for the same $\tau$ and  porous media. Only when $\tau-0.5$ is close to zero, they approach to their intrinsic permeability. Note, slightly different porosity will lead to very insignificant change to the intrinsic permeabilities. But the significantly different normalised permeabilities indicate that great care is required to choose right resolution and $\tau$. If we re-plot the normalized permeability against the Knudsen number, which is $Kn=\sqrt{\pi/6}(\tau-0.5)/N_x$ \cite{Zhangetal2005} for the D2Q9 and D3Q19 lattice models \cite{Qianetal1992}, we find that the results also collapse into a single line especially for small {\it Kn}, see Fig.~\ref{fig2d:Kn}. From Fig.~\ref{fig2d:Kn}, we can clearly see the choice of parameters $\tau$ and $N_x$ should ensure that $Kn<10^{-3}$ to simulate intrinsic permeability. Here, $Kn$ is the global value based on the whole length of the domain i.e. $L$, and the local Knudsen number can be much larger to invalid Navier-Stokes level simulation even at $Kn=0.01$. For a large global $Kn$, i.e. $Kn>10^{-3}$ here, D2Q9 model is not accurate enough to capture rarefaction effects. Although the choice of parameter range may be wider for multi-relaxation model, the underlying mechanism is still the same, i.e. LBM simulations of intrinsic permeability should choose appropriate $\tau$ and resolution i.e. $N_x$ to ensure $Kn$ is small to exclude rarefaction effect.

\begin{figure}[H]
\centering
  \includegraphics[width=0.6\textwidth]{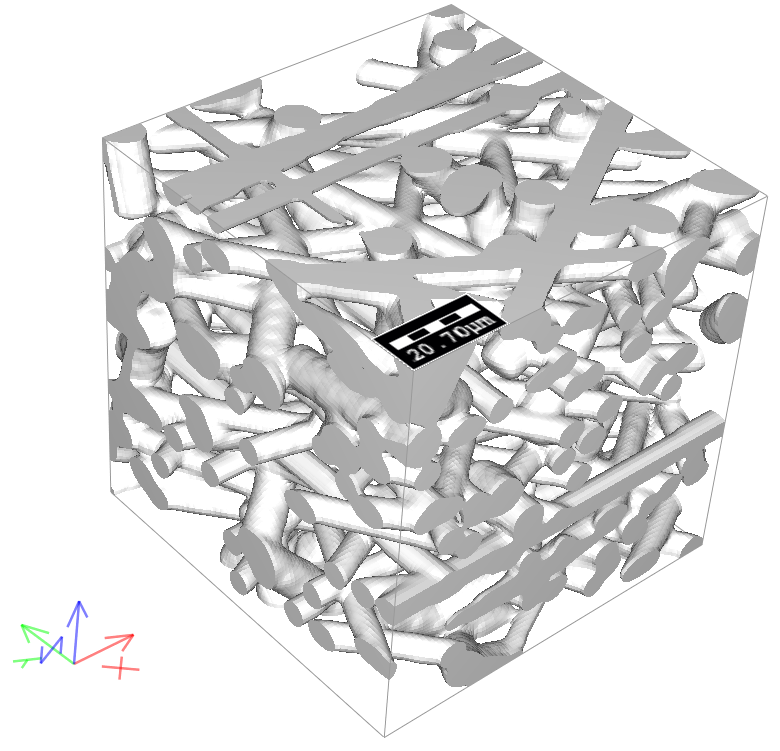}
\caption{The geometry used in the 3D pore-scale simulations.}
\label{fig:geometry}
\end{figure}

We now use LBM to simulate a force-driven flow in a more realistic 3D porous media, see Fig.~\ref{fig:geometry}, as another example to study intrinsic permeability. The flows here are always driven in the $x$ direction, and the volumetric velocity $\left<\vec u\right>_\Omega$ is used to compute the permeability $\vec\kappa$ m$^2$ according to the Darcy law when an external force $\vec g=(g,0,0)$ m/s$^2$ is applied:

\begin{equation} \label{eq:kappa}
\begin{aligned}
   \vec\kappa&=\dfrac{\nu\left<\vec u\right>_\Omega}{g}=\dfrac{\nu}{g}\dfrac{\sum_{k\in\rm{fluid}}\vec u_k}{N_{\rm grid}},
\end{aligned}
\end{equation}
where $\sum_{k\in\rm{fluid}}\vec u_k$ is summation of the flow velocity $\vec u_k$ over all fluid grids $k$, and $N_{\rm grid}$ is the total grid number. We only show the permeability component in the driven direction in the following discussions the same as the above 2D case.

To calculate permeabilities of two similar geometries having the same $N_{\rm grid}$ but $\Delta x_2=0.1\Delta x_1$, we can select model parameters based on Eq.~\eqref{eq:nu} to ensure both the Froude number (\textit{Fr}) and Reynolds number (\textit{Re}) are the same, so that the calculated distribution functions will be the same. In these two sets of parameters, one choice is to have $\Delta t_2=\Delta t_1$ {(so $c_2=0.1c_1$)}, $\tau_2=\tau_1$ (so $\nu_2=0.01\nu_1$ according to Eq.~\eqref{eq:nu}), and $g_2=0.1g_1$. Thus, at the Navier-Stokes level, the velocity solutions of two cases satisfy $\vec u_2=0.1\vec u_1$ and we have $\vec\kappa_2=0.01\vec\kappa_1$ according to Eq.~\eqref{eq:kappa}. Obviously, it is consistent with the following normalized incompressible Navier-Stokes equation:

\begin{equation} \label{eq:normalizedN-S}
\begin{aligned}
   \dfrac{\partial\vec u^\prime}{\partial t^\prime}+\vec u^\prime\cdot\dfrac{\partial\vec u^\prime}{\partial \vec x^\prime}=-\dfrac{\partial(\dfrac{p}{\rho U^2})}{\partial \vec x^\prime}+\dfrac{\nu}{UL}\Delta^\prime\vec u^\prime+\dfrac{L\vec g}{U^2},
\end{aligned}
\end{equation}
where $U$ and $L$ are the characteristic velocity and length to define the normalized parameters $t^\prime=tU/L$, $\vec x^\prime=\vec x/L$ and $\vec u^\prime=\vec u/U$. Eq.~\eqref{eq:normalizedN-S} also implies that more choices are possible to produce the same result (i.e. $\vec\kappa_2=0.01\vec\kappa_1$), e.g., $\Delta t_2=0.1\Delta t_1$ (so $c_2=c_1$), $\tau_2=\tau_1$ (so $\nu_2=0.1\nu_1$ according to Eq.~\eqref{eq:nu}), and $g_2=10g_1$. Consequently, we have $\vec u_2=\vec u_1$ at the Navier-Stokes level. This analysis indicates that the same intrinsic permeability should be obtained with different choice of model parameters at the Navier-Stokes level.

However, both $\tau$ and $N_{\rm grid}$ can affect calculated value of permeability as previously reported \cite{Pan2006}-\cite{Prestininzi2016} and demonstrated in the above 2D case. As Fig.~\ref{fig2d:Kn} shows, it is actually through $Kn$, which indicates how far the flow is away from the Navier-Stokes hydrodynamics. So we examine how $Kn$ affect the calculation of intrinsic permeability in this 3D case.

\begin{figure}[H]
	\centering
	\includegraphics[width=0.6\textwidth]{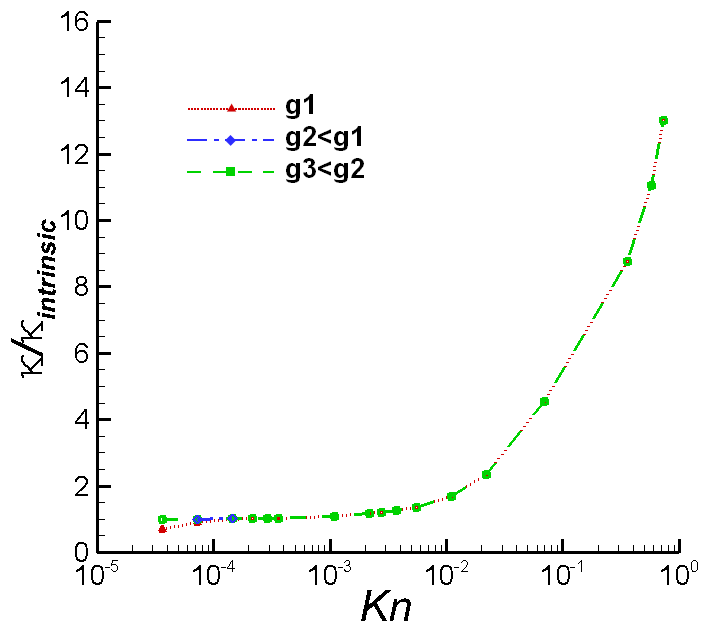}
	\caption{Variation of permeability in the driven direction with $Kn$ for different $g$.}
	\label{fig:3D permeability}
\end{figure}

The computational domain of Fig.~\ref{fig:geometry} has $100^3$ voxels (i.e. grids) and the porosity $\phi$ of $0.748587$. The D3Q19 model is used with six periodic computational boundaries. Fig.~\ref{fig:3D permeability} shows how the permeability changes with $Kn$ for three different values of $g$. We can observe that the permeability is the same for different values of $g$ except at the small $Kn$ for the large $g$, which is due to the increase of inertial effect while the kinematic viscosity decreases with $Kn$ via $\tau$ at a fixed lattice velocity $c$. The permeabilities of the cases with smaller $g$ will also drop if we keep decreasing $Kn$ via $\tau$ leading to a significant increase of $Re$. Therefore, in addition to $Kn$, inertial effect has to be checked in calculation of intrinsic permeability. If we remove the nonlinear velocity terms in equilibrium distribution function, equivalent to solving the Stokes equation i.e. the Navier-Stokes equation without the convective term \cite{LiBrown2013}, the permeability becomes the same for all three values of $g$.

\section{Conclusion}
Since the intrinsic permeability concerned here is defined by the Darcy law for small $Re$, where the viscosity effect is dominant, it is important to choose appropriate model parameters according to Eq.~\eqref{eq:nu} to make sure both $Re$ and $Kn$ are small in computing the intrinsic permeability. Since the resolution is usually low in pore-scale simulations to avoid high computational cost, the choice of $\tau$ becomes more restricted as we need to make sure both $Re$ and $Kn$ are small. If we can afford high resolution, rarefaction effect may be excluded at a large $\tau$ by grid refinement.



\end{document}